\documentclass[aps,prl,preprint,groupedaddress]{revtex4}

\usepackage{graphicx}
\usepackage{booktabs}
\usepackage{amssymb}
\usepackage{amsmath}
\usepackage{color}

\begin{document}
\title{Fast Particle Tracking With Wake Fields}

\author{M. Dohlus, K. Fl\"ottmann, C. Henning}

\begin{abstract}
Tracking calculations of charged particles in electromagnetic fields 
require in principle the simultaneous solution of the equation of motion
and of Maxwell's equations. In many tracking codes a simpler
and more efficient approach is used: external fields like that of the
accelerating structures are provided as field maps, generated in separate
computations and for the calculation of self fields the model of a
particle bunch in uniform motion is used.
We describe how an externally computed wake function can be approximated
by a table of Taylor coefficients and how the wake field kick can be calculated
for the particle distribution in a tracking calculation. The integrated kick,
representing the effect of a distributed structure, is applied at a discrete time.
As an example, we use our approach to calculate the emittance growth of a bunch
in an undulator beam pipe due to resitive wall wake field effects.
\end{abstract}

\maketitle

\section{Introduction}
Self consistent particle tracking needs the simultaneous solution of the
equation of motion for a multi particle system ($N \sim 10^3 \cdots 10^9$) and
of Maxwell's equations. The effort is usually rather high, as the
electro-magnetic field calcualation in time domain has to resolve the
dimension of the bunch and of the surrounding geometry: a large volume has to
be discretized with high resolution, and a good spatial resolution is
related to a fine time step \cite{Schnepp}.

\begin{figure}
\includegraphics[width=1.0\textwidth]{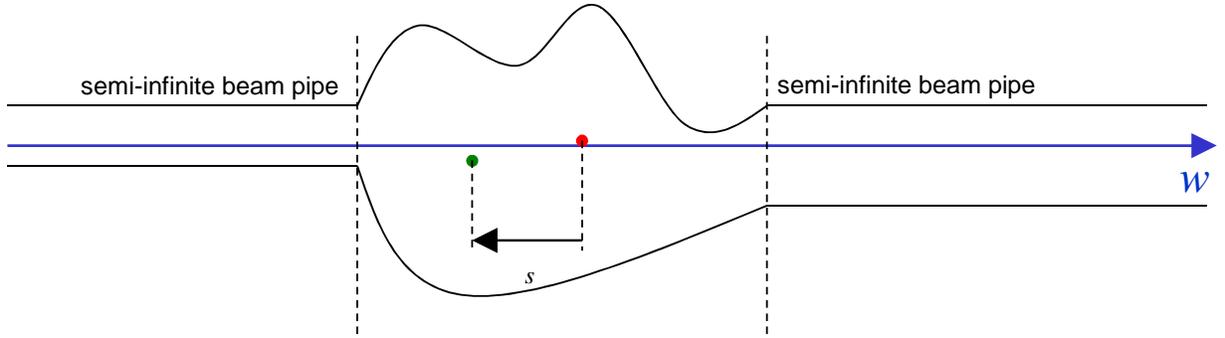}
\caption{Source particle (red) and observer particle (green) in a cavity between semi
infinite beam pipes.}
\label{wake_geometry}
\end{figure}

Supposed the particle energy is high enough for ultra-relativistic
approximations, the motion is approximately uniform and the betatron
wavelength is large compared to geometric dimensions as cavities,
magnets and wake generating discontinuities. Then the electromagnetic
problem can be split into independent sub-problems: external fields, wake field
interactions and simplified self-interactions. The separation of external
fields is possible without these conditions, but the wake field approach
is based on an idealised setup as sketched in Fig.~\ref{wake_geometry}: a
discontinuity, either of geometry or of material properties or of both, is
enclosed by semi-infinite beam pipes. Several generations of wake field codes
have been developed to solve this problem \cite{Weiland,Mafia,Zagorodnov,Henke,Gjonaj}.
The model of simplified self interactions calculates so called ``space charge forces''
for a particle distribution in uniform motion, using a Lorentz transformation
and solving an electrostatic field \cite{Parmela,Floettmann,GPT}.

This report is about the application of wake fields in a tracking program.
The effect of a distributed wakefield generating structure is replaced by a
discrete kick at a discrete time, calculated from a wake function that is
determined with help of a wake field code. The wake function has to be
described by a flexible format. We propose a hybrid formulation: the dependency
on the longitudinal coordinate $s$ (between source and observer particle)
is tabulated and the transverse dependency (offset of source and observer
particle) is Taylor expanded. Therefore the wake function is stored in a
table of Taylor coefficents. The point to point interaction between all
particles is calculated by a convolution method. Therefore the line charge
density and generalized line charge densities (the first and second
transverse moments) are calculated on a grid by a binning and smoothing
technique.

Our method is implemented in the space charge code ASTRA \cite{Floettmann}.
As an example we calculate the emittance growth of a bunch in an undulator
beam pipe due to resitive wall wake field effects. In this example the
wake per length in a 30 m beam pipe is modeled by about 60 diskrete kicks.
The example considers a real focussing lattice, collective and individual
transverse motion and simultaneously longitudinal and transverse effects
(which cannot be separeted from each other).

\section{Coordinate System and Wake Function}
The particles are tracked in global Cartesian coordinates with $(x,y,z,p_x,p_y,p_z)$, the
components of location and momentum, while wake fields are calculation in a local
coodinate system with $(u,v,w,p_u,p_v,p_w)$ components. The coordinates of
both systems are related by:
\begin{eqnarray}
    \left( \begin{array}{ccc} \vec{e}_x & \vec{e}_y & \vec{e}_z \end{array} \right)
    \left( \begin{array}{c} x_n \\ y_n \\ z_n \end{array} \right)
    = 
    \vec{r}_o+
    \left( \begin{array}{ccc} \vec{e}_u & \vec{e}_v & \vec{e}_w \end{array} \right)
    \left( \begin{array}{c} u_n \\ v_n \\ w_n \end{array} \right)
\label{c_tramsform1}
\\
    \left( \begin{array}{ccc} \vec{e}_x & \vec{e}_y & \vec{e}_z \end{array} \right)
    \left( \begin{array}{c} p_{x,n} \\ p_{y,n} \\ p_{z,n} \end{array} \right)
    = 
    \left( \begin{array}{ccc} \vec{e}_u & \vec{e}_v & \vec{e}_w \end{array} \right)
    \left( \begin{array}{c} p_{u,n} \\ p_{v,n} \\ p_{w,n} \end{array} \right)
\label{c_tramsform2}
\end{eqnarray}
with $n$ the particle index, $\vec{r}_o$ the vector to the origin of the wake
geometry, $\vec{e}_w$ the unity vector into nominal direction of particle motion
and $\vec{e}_u$, $\vec{e}_v$ the orthonormal transverse vectors.  The discrete
wake induced kicks are added to the particle momenta when the center of mass of the
distribution passes the plane $\vec{r}_o+u \vec{e}_u +v \vec{e}_w$. The setup is skeched in Fig.~\ref{wake_geometry}.

For simplicity we set $\vec{r}_o=\vec{0}$ for the following. A source particle with
charge $q_s$ on the trajectory
$\vec{r}_s(t)=u_s \vec{e}_u + v_s \vec{e}_v+ ct \vec{e}_w$ is followed by a observer
particle with charge $q_o$ and trajectory
$\vec{r}_o(t)=u_o \vec{e}_u + v_o \vec{e}_v+(ct-s) \vec{e}_w$. The source particle
causes the electromagentic field $\vec{E}^{(s)}$, $\vec{B}^{(s)}$. The total change
of momentum of the observer particle is
\begin{equation}
    \Delta \vec{p} = q_o \int_{-\infty}^{\infty} \left(
                         \vec{E}^{(s)} ( \vec{r}_o(t),t)
                         + c \vec{e}_w \times \vec{B}^{(s)} ( \vec{r}_o(t),t)
                                                 \right) dt
    .
\end{equation}
For the definition of the wake function we follow \cite{Wanzenberg} with
\begin{equation}
    \vec{w}_f(u_s,v_s,u_o,v_o,s)=\frac{c}{q_o q_s} \Delta \vec{p}
    \label{wake_function}
    .
\end{equation}
The wake function is causal in the $s$ coordinate, the transverse and longitudinal
components are related by the Panofsky-Wenzel-Theorem
\begin{eqnarray}
    \frac{\partial}{\partial s} (\vec{w}_f \cdot \vec e_u)
                = - \frac{\partial}{\partial u_o} (\vec{w}_f \cdot \vec e_w)
     \: \:  , \\
    \frac{\partial}{\partial s} (\vec{w}_f \cdot \vec e_v)
                = - \frac{\partial}{\partial v_o} (\vec{w}_f \cdot \vec e_w)
    \label{Panofsky_Wenzel}
     \: \:  ,
\end{eqnarray}
and the longitudinal component is a harmonic function of transverse coordinates
\begin{equation}
    \left( \frac{\partial^2}{\partial u_o^2} +
           \frac{\partial^2}{\partial v_o^2} \right)
       (\vec{w}_f \cdot \vec e_w)=0
    \label{longitudinal_harmonic}
    .
\end{equation}
For the following we use scalar component functions
\begin{equation} 
    \vec{w}_f(u_s,v_s,u_o,v_o,s) = 
    -\left( \begin{array}{ccc} \vec{e}_u & \vec{e}_v & \vec{e}_w \end{array} \right)
    \left( \begin{array}{c} h_u(u_s,v_s,u_o,v_o,s) \\
                            h_v(u_s,v_s,u_o,v_o,s) \\
                            h_w(u_s,v_s,u_o,v_o,s) \end{array} \right)
    .
\end{equation}
The implementation of wake fields in Astra is based on a second order Taylor
expansion of the longitudinal wake in the transverse coordinates
\begin{equation}
    h_w(u_s,v_s,u_o,v_o,s) =
                  \left[ \begin{array}{c} 1 \\ u_s \\ v_s \\ u_o \\ v_o \end{array} \right]^t
                  \left[ \begin{array}{ccccc} h_{00}(s) & h_{01}(s) & h_{02}(s) & h_{03}(s) & h_{04}(s)  \\
                                              0         & h_{11}(s) & h_{12}(s) & h_{13}(s) & h_{14}(s)  \\
                                              0         & h_{12}(s) & h_{22}(s) & h_{23}(s) & h_{24}(s)  \\
                                              0         & h_{13}(s) & h_{23}(s) & h_{33}(s) & h_{34}(s)  \\
                                              0         & h_{14}(s) & h_{24}(s) & h_{34}(s) &-h_{33}(s)  
                         \end{array} \right]
                  \left[ \begin{array}{c} 1 \\ u_s \\ v_s \\ u_o \\ v_o \end{array} \right]
    .
\end{equation}
This approach fulfilles Eq.~(\ref{longitudinal_harmonic}). The transverse components are uniquely related to the longitudinale wake by causality and Panofsky-Wenzel-Theorem:
\begin{eqnarray}
         h_u(u_s,v_s,u_o,v_o,s) &=& h_{03}^{(i)}(s)+2h_{13}^{(i)}(s)u_s+2h_{23}^{(i)}(s)v_s+2h_{33}^{(i)}(s)u_o+2h_{34}^{(i)}(s)v_o
   \: \:  ,  \\
         h_v(u_s,v_s,u_o,v_o,s) &=& h_{04}^{(i)}(s)+2h_{14}^{(i)}(s)u_s+2h_{24}^{(i)}(s)v_s+2h_{34}^{(i)}(s)u_o-2h_{33}^{(i)}(s)v_o
   \: \:  ,
\end{eqnarray}
with the integrated coefficient functions
\begin{equation}
         h_{\alpha \beta}^{(i)}(s)=-\int_{-\infty}^s h_{\alpha \beta}(x) dx
   \: \: .
\end{equation}
Special cases for geometries with symmetry of revolution are the monopole
and dipole wake. The monopole wake 
\begin{equation} 
    \vec{w}_f(u_s,v_s,u_o,v_o,s) = -h_{00}(s) \vec{e}_w
\end{equation}
is purely longitudinal and independent on offset parameters. The transverse part
of the dipole wake depends linear on the offset of the source particle. Due to
symmetry the coefficient functions $h_{13}(s)$ and $h_{24}(s)$ are identical.
Therefore the dipole wake functions is
\begin{equation} 
    \vec{w}_f(u_s,v_s,u_o,v_o,s) = -(u_s \vec{e}_u+v_s \vec{e}_v) 2h_{13}^{(i)}(s) 
                                   -(u_s u_o+v_s v_o) \vec{e}_w 2h_{13}(s)
   \: \: .
\end{equation}
\section{Distributed Source}
For an arbitrary distribution of source particles with charges $q_{s,n}$, total
charge $q_b=\sum q_{s,n}$ and trajectories
$\vec{r}_{s,n}(t)=u_{s,n} \vec{e}_u + v_{s,n} \vec{e}_v+ (ct+ w_{s,n}) \vec{e}_w$,
the change of momentum of the observer particle is
\begin{equation}
        \Delta \vec{p} = \frac{q_o}{c} \sum_n q_{s,n} \vec{w}_f(u_{s,n},v_{s,n},u_o,v_o,s-w_{s,n})
   \: \: .
   \label{delta_p_particle_sum}
\end{equation}
The wake potential 
\begin{equation}
    \vec{W}(u_o,v_o,s) = \frac{1}{q_b} \sum_n q_{s,n} \vec{w}_f(u_{s,n},v_{s,n},u_o,v_o,s-w_{s,n})
    .
    \label{wake_potential}
\end{equation}
characterises the shape dependent wake kick. It can be computed with electromagnetic
field calculation programs as \cite{Weiland,Mafia,Gjonaj} for continuous gaussian source distributions.
Usually the wake function is extrapolated from wake potential computations for small
source distributions.

The particle summation in Eq.~(\ref{delta_p_particle_sum}) is replaced by a
volume integration
\begin{equation}
        \Delta \vec{p} = \frac{q_o}{c} \int 
                                           \varrho(\tilde{u},\tilde{v},\tilde{w})
                                           \vec{w}_f(\tilde{u},\tilde{v},u_o,v_o,s-\tilde{w}) 
                                       d\tilde{u} d\tilde{v} d\tilde{w}
   \label{delta_p_particle_integral}
\end{equation}
with $\varrho(u,v,w)$ the charge density. Formally this can be expressed with a Taylor
expansion of the wake function as
\begin{equation}
        \Delta \vec{p} = \frac{q_o}{c}  
                                       \int 
                                           \varrho(\tilde{u},\tilde{v},\tilde{w})
                 \sum_{ijmn} \left( \vec{w}_{ijmn}(s-\tilde{w}) \tilde{u}^i \tilde{v}^j u_o^m v_o^n \right)
                                       d\tilde{u} d\tilde{v} d\tilde{w}
   \: \: .
\end{equation}
Changing the order of summation and integration $\Delta \vec{p}$ can be written as
sum of one dimensional convolution integrals
\begin{equation}
        \Delta \vec{p} = \frac{q_o}{c} \sum_{ijmn} u_o^m v_o^n
                                       \int 
                                           \lambda_{ij}(\tilde{w})
                                           \vec{w}_{ijmn}(s-\tilde{w}) 
                                       d\tilde{w}
   \: \: , 
   \label{sum_of_convol}
\end{equation}
with scalar functions
\begin{equation}
        \lambda_{ij}(\tilde{w}) =      \int 
                                           \varrho(\tilde{u},\tilde{v},\tilde{w})
                                           \tilde{u}^i \tilde{v}^j 
                                       d\tilde{u} d\tilde{v}
      \: \: ,                                    
   \label{gen_charge_dens}
\end{equation}
and $\vec{w}_{ijmn}(s)$ described in Tab.~\ref{w-expansion}.
$\lambda_{00}(w)$ is the line charge density. In the following we call
$\lambda_{ij}(w)$ generalized line charge densities.
\section{Numerical Realization}
For the numerical kick calculations three problems have to be solved:
the calculation of generalized one dimensional density functions $\lambda_{ij}(s)$,
the representation of coefficient functions  $h_{\alpha \beta}(s)$
and the convolution of these functions.

\subsection{Binning and Smoothing}
A {\sl continuization} based on a binning and smoothing technique is used to
convert the particle set with longitudinal positions $\left\{ w_n \right\}$ and discrete
generalized charges $\left\{ q_{s,n}u_n^i v_n^j \right\}$ to continuous generalized line
charge densities $\lambda_{ij}(s)$. Conventional binning distributes the
particle charges to $N_{bin}$ equi-spaced bins ranging over the whole bunch 
length. The disadvantages are a spacial resolution independent on particle density,
and the bin boundaries are determined by extreme particles (first and last particle).
To achive more flexibility, we use bins that fulfill requirements for the
length per bin as well as for the charge per bin, and we utilize a multiple
binning with $N_{sub}$ different conditions for the first and last bin.

\begin{figure}
\includegraphics[width=1.0\textwidth]{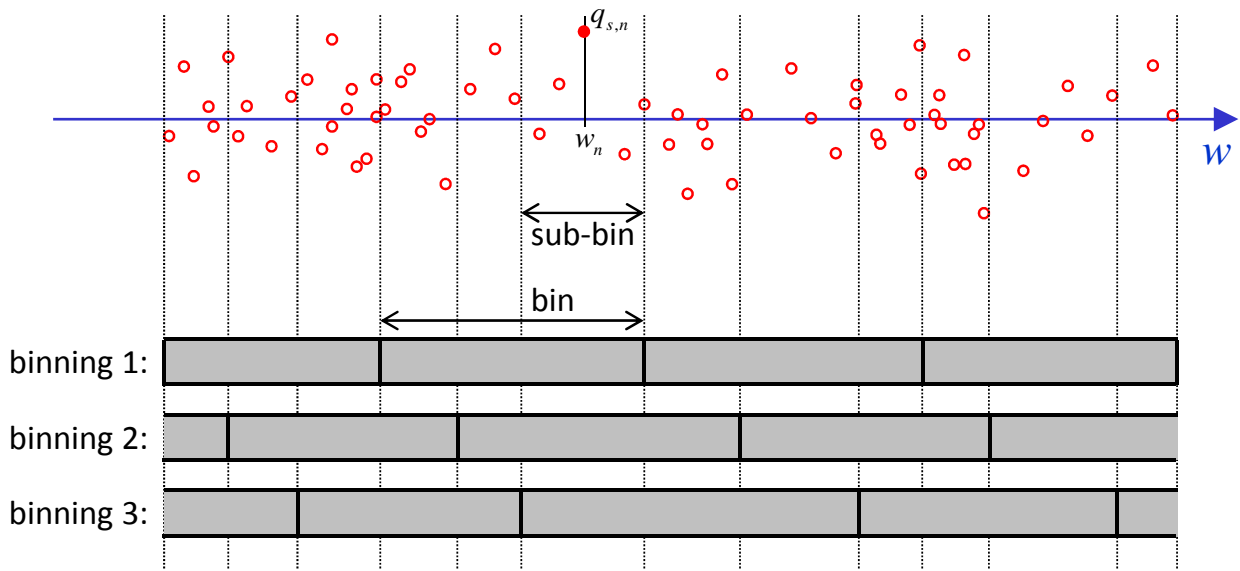}
\caption{Binning and sub-binning for $N_{bin}=4$ and $N_{sub}=3$.}
\label{binning}
\end{figure}

Therefore the bunch is split into $N_{bin} N_{sub}$ sub-bins. All inner bins
are composed by $N_{sub}$ consecutive sub-bins, the outer bins may combine 
less sub-bins, as sketched in Fig.~\ref{binning}. By doing this, $N_{sub}$
different binnings are generated. 

The sub-binning is controlled by the linear combination
$f(w)=W_{eq}f_w(w)+(1-W_{eq})f_q(w)$ of two auxiliary functions
$f_w(w)$ and $f_q(w)$. These functions increase along the bunch monotonously
from zero to one. Function $f_w(w)$ is proportional to length while
$f_q(w)$ increases with the integrated bunch charge. The boundaries $b_k$ of
the sub-bins are the solutions of $f(b_k)=k/(N_{bin} N_{sub})$. Setting
$W_{eq}$ to one causes equi-spaced sub-bins, zero leads to equi-charged
sub-bins and a middle value generates a mixture of both conditions. 
The same bins and sub-bins are used for all generalized charges.

Result of the binning procedure are $N_{bin}N_{sub}+N_{sub}-1$ different bins, each
characterized by  the bin center $b_{c,k}=(b_k+b_{k+N_{sub}})/2$, the bin width
$b_{w,k}=(b_{k+N_{sub}}-b_k)$ and the bin weights (the sums of generalized charges)
\[
     Q_{ij,k}=\frac{1}{N_{sub}} \sum_{n \in \mbox{bin}_k} q_{s,n}u_n^i v_n^j
      \: \: .                                    
\]

The smoothing procedure replaces all bins by continuous functions with certain centers,
widths and weights and samples the sum of these functions on an equidistant grid:
\begin{eqnarray}
     \lambda_{ij,n}&=&\lambda_{ij}(n\Delta w)
     \label{smooth_gen_line_charge_1} \: \: , \\
     \lambda_{ij}(w)&=&\sum_k \frac{Q_{ij,k}}{b_{w,k}} S(\frac{w-b_{c,k}}{b_{w,k}})
     \label{smooth_gen_line_charge_2} \: \: . 
\end{eqnarray}
The ASTRA implementation uses rectangular, triangular and gaussian smoothing
functions:
\begin{eqnarray*}
     S_r(x)&=&\left\{ \begin{array}{ll} 1 & |x|<0.5 \\
                                      0 & \mbox{otherwise} \end{array} \right.
     \: \: ,   \\
     S_t(x)&=& \max(0,1-|x|)
     \: \: ,   \\
     S_g(x)&=& \frac{1}{\sqrt{2\pi}p} \exp\left(-\frac{x^2}{2p^2}      \right)
     \: \: ,
\end{eqnarray*}
with $p \sim 1$, a control parameter. The grid density is determined by the smallest bin width.
\subsection{Representation of Coefficient Functions}
\begin{figure}
\includegraphics[width=1.0\textwidth]{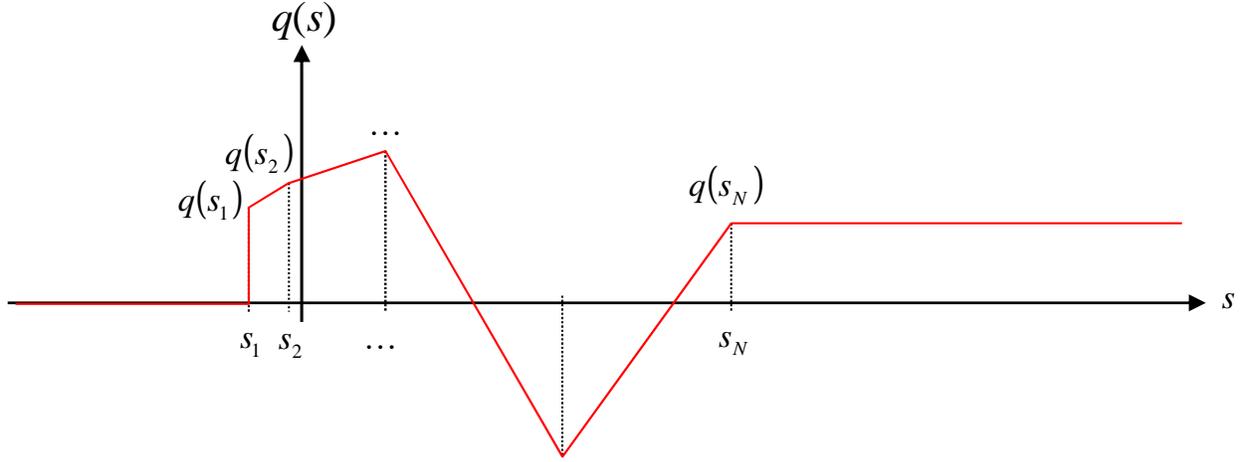}
\caption{Auxiliary function $q(s)$ of poligonial shape, with $s_1 < s_2 < ... < s_N$.}
\label{aux_function}
\end{figure}

Each coefficient function $h_{\alpha\beta}(s)$ is described by
coefficients $R_{\alpha\beta}, L_{\alpha\beta}, \tilde{C}_{\alpha\beta}$ and limited
auxiliary functions $q_{\alpha\beta}(s)$, $p_{\alpha\beta}(s)$.
For the following we skip the index. The representation is
\begin{equation}
    h(s)=q(s)+\frac{\Phi(s)}{C}+Rc\delta(s)-c\frac{\partial}{\partial s}
         \left[ Lc\delta(s)+p(s) \right]
     \: \: ,
     \label{repres_coef}
\end{equation}
with $\delta(s)$ the dirac function, $\Phi(s)$ the step function and $C=\tilde{C}$
if $\tilde{C}\ne 0$. The term with $1/C$ is skipped for $\tilde{C}=0$. The
representation is not unique, but very flexible: every coeffcient $R, L, C$ might
vanish and the auxiliary functions migth be identical to zero. The auxiliary
functions are polygons as sketched in Fig.~\ref{aux_function}. The table description
of auxiliary function $h_{\alpha\beta}(s)$ is listed in Tab.~\ref{table_aux_function}.
The entry "$10\alpha+\beta$" describes the subscript of the auxiliary function.
For a complete wake field representation all sub-tables of non vanishing coefficient
functions are stacked to one table
\[
    \mbox{Table}=\mbox{stack} \left( \left[ \begin{array}{cc} N_t & 0 \end{array} \right] ,
                                     \mbox{Table1}, \mbox{Table2} \ldots \right)
     \: \: ,
\]
with $N_t$ the number of sub-tables.
The order of sub-tables is arbitrary, vanishing coefficients need no representation.

\subsection{Convolution}
According to Eqs.~(\ref{sum_of_convol},\ref{gen_charge_dens})
we need convolution integrals of generalized charges with coefficient functions and
integrated coefficient functions:
\begin{eqnarray*}
    H(s)        &=& \int \lambda(x) h(s-x) dx        \: \: ,\\
    H^{(i)}(s)  &=& \int \lambda(x) h^{(i)}(s-x) dx  \: \: .
\end{eqnarray*}
For simplicity we skipped all lower indices. Using the representation Eq.~(\ref{repres_coef}) and
auxiliary functions
\begin{eqnarray*}
    A(s)&=&q(s)+\Phi(s)/C
  \: \: , \\
    B(s)&=&-cp(s)
  \: \: , \\
    Q(s)&=&\int_{-\infty}^s q(\tilde{s})d\tilde{s}
  \: \: , \\
    C(s)&=&-Q(s)+cp(s)-(s/C+Rc)\Phi(s)
  \: \: ,
\end{eqnarray*}
the convolutions can be rewitten as
\begin{eqnarray}
    H(s)&=&  Rc\lambda(x) - Lc^2\lambda'(x) + \int \lambda(x) A(s-x) dx + \int \lambda'(x) B(s-x) dx
    \: \: , \\
    H^{(i)}(s)&=& Lc^2\lambda(x) + \int \lambda(x) C(s-x) dx
    \: \: .
\end{eqnarray}
The integrals are solved piecewise for the generalized charged densities Eq.~(\ref{smooth_gen_line_charge_2})
and polygonal coeffcient functions.

\section{Example}
The example of particle motion in an undulator vacuum chamber with resisitve wall wake fields
is based on parameters used in \cite{Schlarb}, see Tab.~\ref{schlarb_table}.
For the calculation with ASTRA, a round stainless steel beam pipe with radius $r_b =4.75 $ mm is used to
cause significant transverse effects. The undulator chamber of the real facility (FLASH) is made from
aluminium. For the numerical simulation 63 quadrupoles of the FODO lattice are considered and the resistive
wall wake of the 30 m beam pipe is simulated by 62 discrete wake kicks (each for $\Delta L=L_{FODO}/2$)
applied in the middle of the drifts between the quadrupoles.

\begin{figure}
\includegraphics[width=1.0\textwidth]{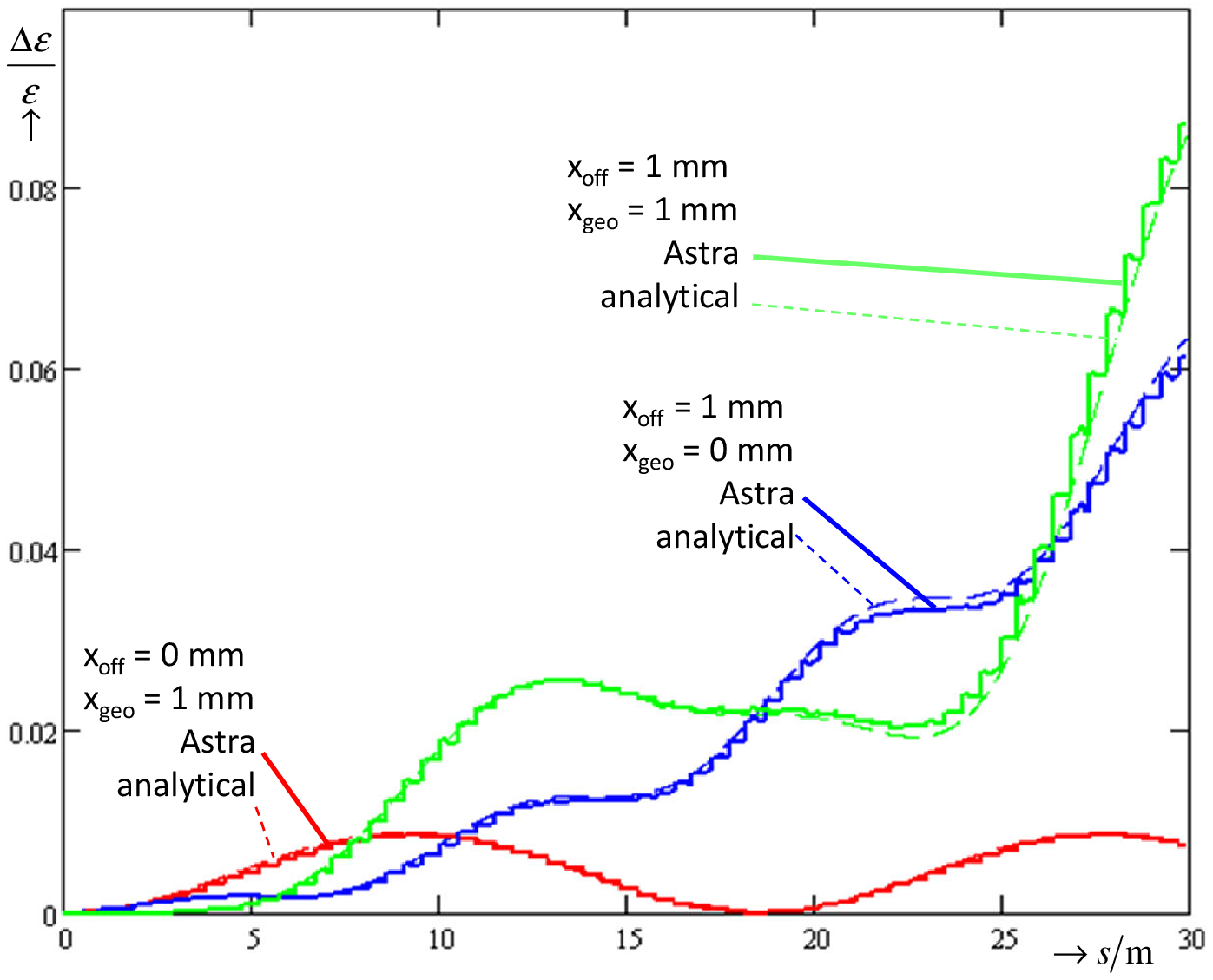}
\caption{Emittance growth along the undulator for dipole wakes in a stainless steel vacuum chamber
with a radius of 4.75 mm: numerical simulation (solid line); analytical solution (dashed line).}
\label{fig4}
\end{figure}

\begin{figure}
\includegraphics[width=1.0\textwidth]{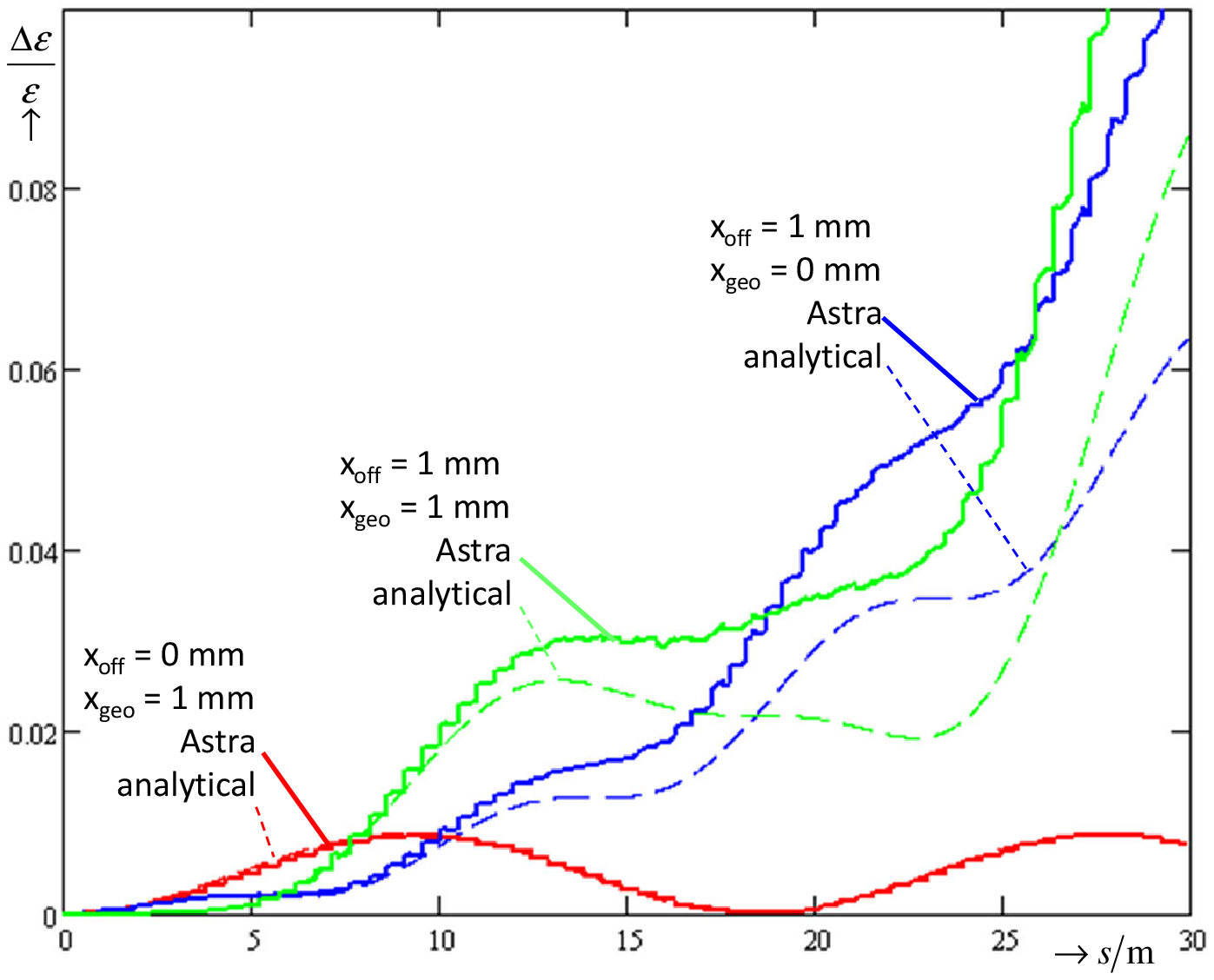}
\caption{Emittance growth along the undulator for monopole and dipole wakes in a stainless steel
vacuum chamber with a radius of 4.75 mm: numerical simulation (solid line); analytical solution
(dashed line).}
\label{fig5}
\end{figure}

The longitudinal wake per length of a round resistive beam pipe is approximated by
\begin{equation}
     \vec{w}_f(u_s,v_s,u_o,v_o,s)\cdot\vec{e_w}=w_r(s) \left(
                                            1+2\frac{u_s u_o + v_s v_o}{r_b^2}
                                                       \right)
     \: \: ,
     \label{long_res_wall_wake}
\end{equation} 
with
\begin{eqnarray*}
     w_r(s)&=&-\frac{1}{2\pi}\int Z_r(\omega) \exp(j\omega s/c)d\omega
   \: \: , \\
     Z_r(\omega)&=&\frac{Z_s(\omega)}{2\pi r_b}\left( 1+j\omega\varepsilon Z_s(\omega) r_b/2  \right)^{-1}
   \: \: , \\
     Z_s(\omega)&=&\sqrt{j\omega \mu / \kappa(\omega)}
   \: \: ,
\end{eqnarray*}
$\mu$ the permeability, $\varepsilon$ the permittivity of vaccum and $\kappa(\omega)$ the
frequency dependent conductivity of the pipe \cite{Bane,Schlarb}. therefore only three coefficient functions
\begin{eqnarray*}
   h_{00}(s)&=&-\Delta L w_r(s)
   \: \: , \\
   h_{13}(s)&=&-\frac{\Delta L w_r(s)}{r_b^2} 
   \: \: , \\
   h_{24}(s)&=&-\frac{\Delta L w_r(s)}{r_b^2} 
   \: \: ,
\end{eqnarray*}
have to be calcualted that are all proportional to $w_r(s)$, the longitudinal resistive wall wake per
length. In the wake table the coefficients $N_p$, $R$, $L$ und $\tilde{C}$ are set to zero
(for $\alpha\beta=00$, $13$ and $24$), only the auxiliary functions  $u_{00}(s)$, $u_{13}(s)$ and $u_{24}(s)$ are used. In Figs.~\ref{fig4} the emittance growth due to dipole wake forces
is shown along the undulator and compared with the analytic estimation from \cite{Schlarb}.
The numerical result agrees well with the estimation, but the dipole model is incomplete.
Fig.~\ref{fig5} shows the emittance growth due to monopole and dipole wake fields. The longitudinal
wake causes an energy spread of abaut 2.4 MeV after 30 meters. For off axis particles with that
energy spread, the quadrupole focussing is significantly altered and the transverse emittance
is further increased.

\section{Acknowledgements}
We thank Torsten Limberg, Rainer Wanzenberg and Igor Zagorodnov for useful discussions and
comments on this work.

%
%

\clearpage
\begin{table}
\begin{tabular}{|l|l|l|l|c|c|c|}                                     \hline
    i & j & m & n & $\vec{w}_{ijmn}(s)\cdot\vec{e}_{u}$  
                  & $\vec{w}_{ijmn}(s)\cdot\vec{e}_{v}$
                  & $\vec{w}_{ijmn}(s)\cdot\vec{e}_{w}$ \\ \hline \hline
    0 & 0 & 0 & 0 &  $-h_{03}^{(i)}(s)$ &  $-h_{04}^{(i)}(s)$ & $ -h_{00}(s) $     \\ \hline
    1 & 0 & 0 & 0 & $-2h_{13}^{(i)}(s)$ & $-2h_{14}^{(i)}(s)$ & $ -h_{01}(s)$     \\ \hline
    0 & 1 & 0 & 0 & $-2h_{23}^{(i)}(s)$ & $-2h_{24}^{(i)}(s)$ & $ -h_{02}(s)$     \\ \hline
    0 & 0 & 1 & 0 & $-2h_{33}^{(i)}(s)$ & $-2h_{34}^{(i)}(s)$ & $ -h_{03}(s)$     \\ \hline
    0 & 0 & 0 & 1 & $-2h_{34}^{(i)}(s)$ & $-2h_{44}^{(i)}(s)$ & $ -h_{04}(s)$     \\ \hline
    2 & 0 & 0 & 0 & $0$               & $0$                   & $ -h_{11}(s)$     \\ \hline
    1 & 1 & 0 & 0 & $0$               & $0$                   & $-2h_{12}(s)$     \\ \hline
    1 & 0 & 1 & 0 & $0$               & $0$                   & $-2h_{13}(s)$     \\ \hline
    1 & 0 & 0 & 1 & $0$               & $0$                   & $-2h_{14}(s)$     \\ \hline
    0 & 2 & 0 & 0 & $0$               & $0$                   & $ -h_{22}(s)$     \\ \hline
    0 & 1 & 1 & 0 & $0$               & $0$                   & $-2h_{23}(s)$     \\ \hline
    0 & 1 & 0 & 1 & $0$               & $0$                   & $-2h_{24}(s)$     \\ \hline
    0 & 0 & 2 & 0 & $0$               & $0$                   & $ -h_{33}(s)$     \\ \hline
    0 & 0 & 1 & 1 & $0$               & $0$                   & $-2h_{34}(s)$     \\ \hline
    0 & 0 & 0 & 2 & $0$               & $0$                   & $  h_{33}(s)$     \\ \hline
\end{tabular}
\caption{$\vec{w}_{ijmn}(s)$ function.}
\label{w-expansion}
\end{table}
\begin{table}
\begin{tabular}{|cl|cl|}                                                      \hline
    $N_q$             & \hspace{1.5cm} &  $N_p$                 &  \hspace{1.5cm}   \\ \hline
    $R$               & /(Us)   &  $L$                          & /(Us$^2$)\\ \hline
    $\tilde{C}$       & /(1/U)  &  $10 \alpha + \beta$          &          \\ \hline
    $s_{q,1}$         & /m      &  $q\left( s_{q,1} \right)$    & /U       \\ \hline
    $s_{q,2}$         & /m      &  $q\left( s_{q,2} \right)$    & /U       \\ \hline
    ...               &         &  ...                          &          \\ \hline
    $s_{q,N_q}$       & /m      &  $q\left( s_{q,N_q} \right)$  & /U       \\ \hline
    $s_{p,1}$         & /m      &  $p\left( s_{p,1} \right)$    & /(Us)    \\ \hline
    $s_{p,2}$         & /m      &  $p\left( s_{p,2} \right)$    & /(Us)    \\ \hline
    ...               &         &  ...                          & /(Us)    \\ \hline
    $s_{p,N_q}$       & /m      &  $p\left( s_{p,N_p} \right)$  & /(Us)    \\ \hline
\end{tabular}
\caption{Sub-table of auxiliary function $h_{\alpha\beta}(s)$. The unit U is V/As for
         $\alpha\beta=00$, V/(Asm) for $\alpha\beta=01\cdots04$ and V/(Asm$^2$) for all
         other coefficients.}
\label{table_aux_function}
\end{table}
\begin{table}
\begin{tabular}{|lc|l|l|}                                     \hline
    variable            &                   &  units      & values  \\ \hline \hline
    beam energy         & $E_0$             &  GeV        & 1       \\ \hline
    bunch charge        & $Q_0$             &  nC         & 1       \\ \hline
    rms bunch length    & $\sigma_z$        &  $\mu$m     & 50.0    \\ \hline
    normalized emittance in the undulator
                      & $\epsilon^n$        &  $\mu$m     & 2.0     \\ \hline
    undulator length    & $L$               &  m          & 30.0    \\ \hline
    undulator gap       & $g$               &  mm         & 12.0    \\ \hline
    pipe thickness (minimum)  
                        & $t$               &  mm         &  1.25   \\ \hline \hline
    beam optics beta function
                        & $\bar{\beta}$     &  m          & 3      \\ \hline
    FODO period length  & $L_{FODO}$        &  m          & 0.96   \\ \hline
    Length of FODO quad & $L_{FODO}$        &  mm         & 136.5  \\ \hline

\end{tabular}
\caption{Parameters for wake field calculation in undulator as asumed for TTF2.}
\label{schlarb_table}
\end{table}

\clearpage



\begin{thebibliography}{99} \vspace{-2mm}

\bibitem{Schnepp}
  S.~Schnepp, E.~Gjonaj, T.~Weiland: On the Development of a Self-Consistent Particle-In-Cell (PIC)
  Code Using a Time-Adaptive Mesh Technique. Proceedings of the 10th European Particle Accelerator
  Conference (EPAC 2006) (EPAC 2006), June 01.07.2006, pp. 2182-2184
\bibitem{Weiland}
  T.~Weiland: Comment on Wakefield Computation in the Time Domain (TBCI). NIM-A 216, pp. 31-34, 1983.
\bibitem{Mafia}
  MAFIA Collaboration, MAFIA manual, CST GmbH, Darmstadt, 1997.
\bibitem{Zagorodnov}
  I.~Zagorodnov: Indirect methods for wake potential integration. Phys. Rev. ST Accel. Beams 9, 2006.
\bibitem{Henke}
  H.~Henke and W.~Bruns, in Proceedings of EPAC 2006, Edinburgh, Scotland (WEPCH110, 2006).
\bibitem{Gjonaj}
  E.~Gjonaj, T.~Lau, T.~Weiland: Computation of Short Range Wake Fields with PBCI. ICFA Beam Dynamics
  Newsletter (ICFA 2008), April 01.04.2008, pp. 38-52
\bibitem{Wanzenberg}
  T.~Weiland, R.~Wanzenberg: Wake Fields and Impedances, DESY M-91-06.
\bibitem{Parmela}
  L.M.~Young: ``PARMELA'' Los Alamos National Laboratory report LA-UR-96-1835
  (Revised April 22, 2003).
\bibitem{Floettmann}
  K.~Fl\"ottman, ``ASTRA'', DESY, Hamburg, \path|www.desy.de/~mpyflo|, 2000.

\bibitem{GPT}  
  S.~van der Geer, O.~Luiten, M.~de Loos, G.~P\"oplau, U. van Rienen: 3D space-charge model
  for GPT simulations of high brightness electron bunches, Institute of Physics Conference Series, No. 175,
  (2005), p. 101.
\bibitem{Schlarb}
  H.~Schlarb, Resistive Wall Wake Fields, Diploma Thesis 1997.
\bibitem{Bane}
  K.~Bane: The Short-Range Resistive Wall Wakefields. SLAC-PUB-95-7074, Dec. 1995.

\end{thebibliography}
\end{document}